\documentclass[doublecol,figures]{epl2}

\title{Residual entropy in a model for the unfolding of single
polymer chains}
\author{E. Van der Straeten\thanks{Research Assistant of the Research
Foundation - Flanders (FWO - Vlaanderen)}\thanks{\email{Erik.VanderStraeten@ua.ac.be}}\and J. Naudts\thanks{\email{Jan.Naudts@ua.ac.be}}}
\shortauthor{E. Van der Straeten \etal}
\institute{                    
  \inst{1} Departement Fysica, Universiteit Antwerpen, Groenenborgerlaan 171,
2020 Antwerpen, Belgium
}

\abstract{
We study the unfolding of a single polymer chain due to an external force. We
use a simplified model which allows to perform all calculations in closed form without assuming a Boltzmann-Gibbs form for the equilibrium distribution. Temperature is then defined by calcula\-ting the Legendre transform of the entropy under certain constraints. The application of the model is limited to flexible polymers. It exhibits a gradual transition from compact globule to rod. The boundary line between these two phases shows reentrant behavior. This behavior is explained by the presence of residual entropy.
}

\begin{document}

\maketitle

\section{Introduction}
The unfolding of polymers has been studied for many years. Recently, it became
possible to investigate the stability of a single polymer against unfolding by
applying a mechanical force to the end-point of the molecule.
Tools like optical tweezers, atomic force microscopes and
soft microneedles are used in this kind of experiments. The unfolding transition
of polymers is a single- or a multi-step process, depending on the experimental
conditions. For example, with a force-clamp apparatus one is able to study the mechanical
unfolding at constant force. One applies a sudden force which is then kept
constant with feedback techniques. Numerical simulations \cite{referee1} show that a certain protein (called ubiquitin) unfolds in a single step, while an other protein (called integrin) unfolds in multiple steps. Ubiquitin is studied experimentally in \cite{referee2}. The findings of \cite{referee2} support the numerical simulations rather well, because in $95\%$ of the cases a clear single-step unfolding process is observed. In \cite{referee3,referee4}, force-extension relations of single DNA molecules are
obtained in the fixed-stretch ensemble. One measures the average applied force
while keeping the extension of the DNA molecule constant. Depending on the
solvent conditions, force plateaus or stick-release patterns are observed. 

The unfolding transition from compact globule to rod was already
theoretically predicted 15 years ago \cite{referee5}, based on heuristic
arguments. Nowadays, the theoretical study of this transition is dominated by
numerical simulations. The self-avoiding walk (SAW) in two dimensions is
intensively used to model the unfolding transition of polymers
\cite{referee6,referee7,referee8}. The advantage of SAWs is that local
interactions like monomer-monomer attraction and the excluded volume effect are
taken into account. The disadvantage is that one is limited to short walks due
to the computational cost (up to chain length $55$ \cite{referee7}). In
\cite{referee6,referee7,referee8}, a force-temperature state diagram for
flexible polymers in a poor solvent is calculated in the fixed-force ensemble.
The most important property is that the force, at which the polymer unfolds,
goes through a maximum as a function of the temperature. This so called
reentrant behavior is explained by the presence of residual entropy
\cite{referee6}. 

Reentrant behavior has been observed at other occasions as well. A realistic, analytical solvable model for the unfolding transition is presented in \cite{referee8b}. The authors obtain state diagrams for $4$ different molecules. One of these diagrams shows reentrant behavior. The authors of \cite{referee8b} do not comment on this interesting feature. In \cite{referee8bis}, simplified, analytical solvable, lattice models are considered to describe the unzipping of DNA. The obtained state diagram shows reentrant behaviour. This is also caused by the appearance of residual entropy in the model.

In \cite{referee9}, the present authors have proposed a simple model to describe
single polymers. This model has been used in \cite{referee10} to compare the
outcome of two experiments, which are performed in the fixed-force ensemble and the fixed-extension ensemble. In the present letter, we focus on the fixed-force ensemble, because the model is completely solvable in closed form in this ensemble \cite{referee10}. The more general theory of \cite{referee11} allows to simplify the calculations of \cite{referee9,referee10}. Of course, this simplified model is of a qualitative nature. More sophisticated models are needed to describe all the experimentally observed features of the unfolding transition. The aim of the present letter is to focus on one of these features, namely the consequences of the appearance of residual entropy. 

\section{One dimensional random walk with memory}
\begin{figure}
\onefigure[width=0.45\textwidth]{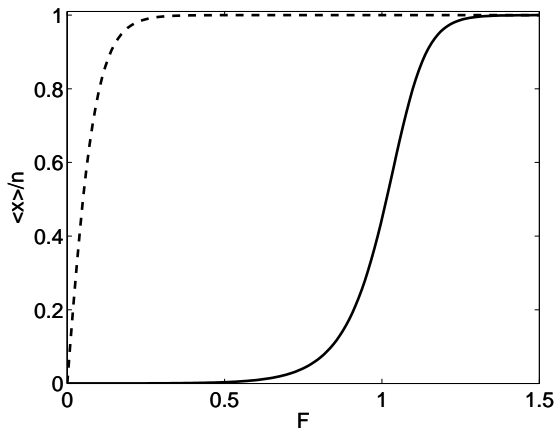}
\caption{Plot of the force-extension relation relation
at constant temperature, $T=0.1$. The value of the parameter $h$ is $-1$ for the solid
line and $0.01$ for the dotted line. The value of $a$ is equal to $1$ for both lines.}
\label{forceexten}
\end{figure}
\begin{figure}
\onefigure[width=0.45\textwidth]{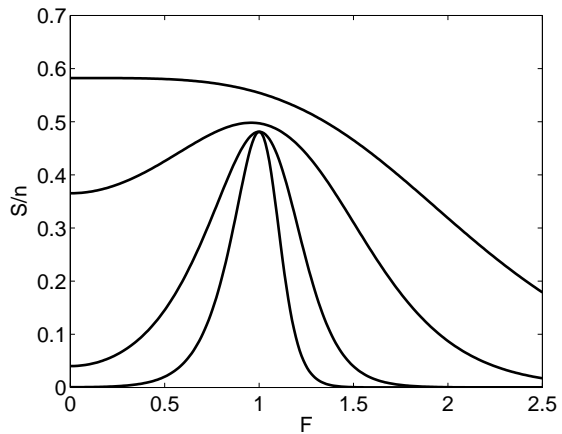}
\caption{Plot of the force-entropy relation at
constant temperature. The values of the temperature are from top till bottom 
$1;0.5;0.2;0.1$. The value of the parameters $h$ and $a$ are equal to $-1$ and $1$ respectively.}
\label{forceentro}
\end{figure}
The model under study is a discrete-time random walk on a one-dimensional
lattice. It depends on two parameters $\epsilon$ and $\mu$, the probabilities to
go straight on, when walking to the right, respectively to the left. This is not
a Markov chain since the walk remembers the direction it comes from. The process
of the increments is a Markov chain with two states, $\rightarrow$ and
$\leftarrow$. The stationary probability distributions of these two states are
\cite{referee9}
\begin{eqnarray}
p(\rightarrow)=\frac{1-\mu}{2-\epsilon-\mu}\ \ \textrm{and}\ \ p(\leftarrow)=\frac{1-\epsilon}{2-\epsilon-\mu}.
\end{eqnarray}
With these expressions, the average end postion $\langle x\rangle$ of the walk and the average number of reversals of direction $\langle K\rangle$ of the walk can be written conveniently as \cite{referee10,referee11}
\begin{eqnarray}\label{gem}
\frac{\langle
x\rangle}{n}&=&a\left[p(\rightarrow)-p(\leftarrow)\right],
\cr
\frac{\langle K\rangle}{n}&=&2(1-\epsilon)p(\rightarrow)=2(1-\mu)p(\leftarrow),
\end{eqnarray}
with $a$ the lattice parameter and $n$ the total number of steps. Also, an expression for the entropy can be obtained in closed form (neglecting boundary terms) \cite{referee10,referee11}
\begin{eqnarray}
\frac Sn&=&p(\rightarrow)\left[-\epsilon\ln\epsilon-(1-\epsilon)\ln(1-\epsilon)\right
]
\cr
&&+p(\leftarrow)\left[-\mu\ln\mu-(1-\mu)\ln(1-\mu)\right].
\end{eqnarray}
We use units in which $k_\textrm B=1$.

This one-dimensional random walk can be used as a simple
model of a flexible polymer in the fixed-force ensemble. The macroscopic
observables are the position of the end point and the number of reversals of
direction (kinks) of the walk. The position of the end point measures the effect
of an external force applied to the end point. The groundstate of a flexible
polymer in a poor solvent is a compact globule.  An obvious definition of the
Hamiltonian is then $H=hK$, with $h$ a negative constant with dimensions of
energy and $K$ the total number of kinks. A positive value of the parameter $h$ correponds with a polymer is a good solvent. The strength of the solution determines the absolute value of $h$. The contour length of the polymer is equal to $na$.

\section{Thermodynamics}
The Legendre transform of $S$ is the free energy $G$
\begin{eqnarray}\label{Legen}
G&=&\inf_{\epsilon,\mu}\left\{E-F\langle
x\rangle-\frac{1}{\beta}S\right\}.
\end{eqnarray}
The solution of the set of equations $\partial G/\partial\epsilon=0$ and
$\partial G/\partial\mu=0$ gives relations for $\beta$ and $F$ as a function of
the model parameters 
\begin{eqnarray}\label{connection_model_therm}
aF=\frac{1}{2\beta}\ln\frac\epsilon\mu&\textrm{and}&\beta=\frac{1}{2h}\ln\frac{
\epsilon\mu}{(1-\epsilon)(1-\mu)}.
\end{eqnarray}
The most general case when $\epsilon\neq\mu$ corresponds with a persistent
random walk with drift. A persistent random walk \cite{referee10bis} without drift is obtained with
the choice $\epsilon=\mu$. This implies $F=0$ but non-vanishing $\beta$. Also
non-persistent
random walk with drift is a special case. This corresponds with $\epsilon+\mu=1$
and implies
$\beta=0$ but non-vanishing $F$. Simple random walk is obtained with
$\epsilon=\mu=1/2$. In this case both $\beta$ and $F$ equal zero. 

The set of equations (\ref{connection_model_therm}) can be inverted in closed
form and has a unique solution for every value of $\beta$ and $F$. We also
calculated the eigenvalues of the matrix of the second derivatives of the free
energy. These eigenvalues are always non-negative. We conclude that the present
model exhibits no phase transition. It is well known that no true phase
transition can occur, because of the finite size of single molecules. This
poses the problem of defining the different phases of the model. In
\cite{referee6}, the sudden change of an appropriate average value is used to
obtain the boundaries between the different phases in the state diagram. We will
use the same criterion to define the boundary line between the different phases
of the present model.
\begin{figure}
\onefigure[width=0.45\textwidth]{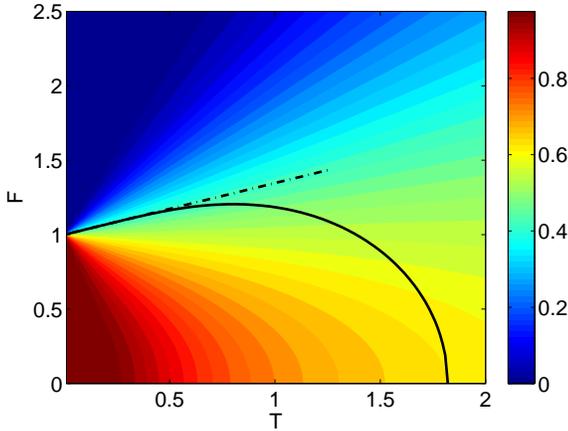}
\caption{(color online). Plot of the average number of kinks
as a function of the temperature and the force. The
color code is mentioned to the right. The black solid line, marks the gradual transition from the compact phase to the stretched phases. The black dotted line is an approximation for the solid
line, valid at low temperatures only. The value of the parameters $h$ and $a$ are equal to $-1$ and $1$ respectively.}
\label{phasekink}
\end{figure}
\begin{figure}
\onefigure[width=0.45\textwidth]{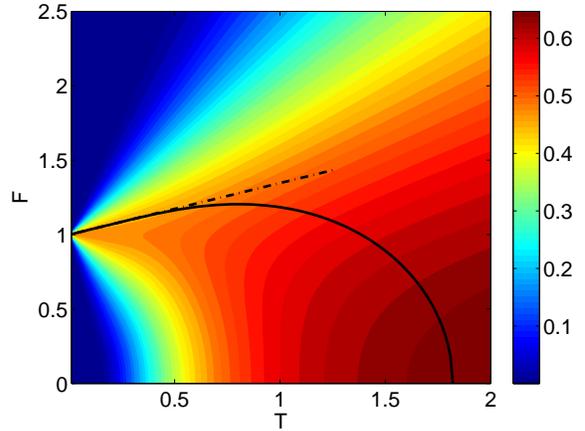}
\caption{(color online). Plot of the entropy as a function of the temperature and the force. The
color code is mentioned to the right. The black solid line, marks the gradual transition from the compact phase to the stretched phases. The black dotted line is an approximation for the solid
line, valid at low temperatures only. The value of the parameters $h$ and $a$ are equal to $-1$ and $1$ respectively.}
\label{phaseentro}
\end{figure}

With expressions (\ref{gem}) and (\ref{connection_model_therm}) one can calculate the force-extension relation at constant temperature. It is shown in figure \ref{forceexten} at low temperature and for two different values of the parameter $h$. For polymers in a bad solvent ($h<0$), one observes a steep increase of the average end-to-end distance at $F\approx1$. For polymers in a good solvent, this steep increase occurs at vanishing force. This in qualitative agreement with experimental observations \cite{referee3,referee4}. In the present letter we will focus on polymers in a bad solvent. At $F\approx1$ the shape of the polymer changes from compact globule to rod. This
becomes a real phase transition (a true force plateau) for $T\rightarrow0$ only.
The smallest eigenvalue of the matrix of the second derivatives of the free
energy vanishes at this moment. The steep increase of the average end-to-end distance at $F\approx1$ disappears for higher
temperatures. Also the force-entropy relation at constant temperature can be
obtained. It is shown in figure \ref{forceentro} for different values of the
temperature. At low temperatures, the entropy goes through a sharp maximum with
increasing force. This maximum disappears at higher temperatures and the entropy
becomes a monotonic decreasing function of the force.

In the present model, the sudden change of the average end-to-end distance can
be used to define the gradual transition from compact globule to rod. The boundary line is obtained from the peak value of $\partial\langle x\rangle/\partial F$ at constant temperature. Figures \ref{phasekink} and \ref{phaseentro} show the average number of kinks and the entropy as a function of force and temperature. The black solid line shows the boundary between the compact phase and the stretched
phases. The end points of the boundary line are $(T_b=0,F_b=-h/a)$ and
$(T_b=-2h/\ln3,F_b=0)$. At low temperatures the boundary line is an increasing
function of the temperature. At intermediate temperatures the boundary line is a
decreasing function of the temperature. In the appendix an
approximated expression for the boundary line at low temperatures is calculated.
One obtains $F=1+0.35T+\ldots$. Figures \ref{phasekink} and \ref{phaseentro} show  the result of the
latter expression together with the exact boundary line. The two lines coincide
up to a temperature of approximately $0.5$. 

In the appendix, we show that the positive slope of the boundary line
at low temperatures is due to the entropy. This is unexpected, because the
entropy is usually not important at low temperatures. However, the entropy of
the present model contains two contributions, the usual thermal entropy and a
configurational entropy, closely related to the zero-temperature phase
transition at $F=1$. It's clear from figures \ref{forceentro} and
\ref{phasekink}, that there are three separate phases at low temperatures. The
globular phase ($F<1$ and $\langle K\rangle/n\approx1$) and the stretched phase
($F>1$ and $\langle K\rangle\approx0$) have a small value of the residual
entropy, because only a limited amount of configurations are allowed. The
intermediate phase at $F\approx1$ has $\langle K\rangle/n\approx 1/2$. This
means there are plenty of allowed configurations. The residual entropy of this
phase is very high. The reentrance of the phase boundary is the consequence of a
very subtle asymmetry in the residual entropy, which prefers the globular phase
above the stretched phase. This asymmetry can most clearly be seen in figure
\ref{forceentro}.

\section{Discussion}
To summarize, we solve a simplified model for the unfolding of a polymer in
closed form in the fixed-force ensemble. The force-extension relation shows an
approximate force plateau at low temperatures. The boundary line between the
globular and stretched phases shows reentrant behavior. This reentrant behavior
is explained by the presence of residual entropy in the model.

We want to stress that the present approach to introduce the temperature deviates from the standard way of introducing temperature in statistical mechanics. Usually, one assumes that the equilibrium probability distribution is of the Boltzmann-Gibbs form $e^{-\beta H}$, with $\beta$ the inverse temperature. The present approach is different. We start from a two-parameter model and calculate the average of the macroscopic variables of interest as a function of these two parameters, without assuming the Boltzmann-Gibbs distribution. Then we define the temperature by calculating the Legendre transform (\ref{Legen}) of the entropy. Usually, the infimum is taken over all possible probability distribution. This results in the Boltzmann-Gibbs distribution. We take the infimum only over the model parameters. This adds an extra constraint on the equilibrium probability distribution. As a consequence, the resulting distribution is not necessarily of the Boltzmann-Gibbs form, although we started from the Boltzmann-Gibbs definition for the entropy \cite{referee11}. Indeed, we already pointed out in \cite{referee9} that the joint probability distribution that after $n$ steps the walk is in $x$ and changed its direction $K$ times is not of the Boltzmann-Gibbs form. The deviations from the Boltzmann-Gibbs form are small and disappear for long chains.

The results of the present paper are limited to the fixed-force ensemble,
although most experiments are performed in the fixed-stretch ensemble. In
\cite{referee10} we show that it is possible to extend the present model to the
fixed-stretch ensemble. However, in this ensemble almost all calculations have
to be performed numerically. We also show in \cite{referee10} that the
differences between the two ensembles vanish in the thermodynamic limit
and are negligible for long chains. So we expect only small, finite size
corrections to the state diagram after extending the present calculations to the
fixed-stretch ensemble. In \cite{referee3,referee4} force plateaus are
experimentally observed in stretching experiments in the fixed-stretch ensemble.
Following the previous reasoning, this is in agreement with the present
calculations up to finite size corrections. To the best of our knowledge, the
temperature dependence of this plateau has not yet been studied experimentally.

As mentioned in the introduction, SAWs are intensively used to study the
unfolding transition of polymers. Reentrant behavior is also observed in these
models \cite{referee6,referee7,referee8}. Starting from a phenomenological
expression for the free energy near $T=0$, one obtains the following boundary
line for flexible polymers
\cite{referee6,referee12}
\begin{eqnarray}\label{fsaw}
F&=&-\alpha+\frac\alpha{\sqrt n}+S_cT,
\end{eqnarray}
with $\alpha$ a negative model parameter. The second term is a surface
correction term. The last term is a contribution due to the residual entropy of
the globular phase, with $S_c$ the entropy per monomer. The presence of
the residual entropy causes the reentrant behavior. The present model does
not contain the surface correction term, because local interactions are not
included. Formula (\ref{fsaw}) is similar to expression (\ref{appr_Fb}) in the
thermodynamic limit, because the surface correction term disappears in this
limit. A simplified version of the SAW is the partially directed walk (PDSAW).
This means that steps with negative projection along the $x$ axis are forbidden.
Analytical calculations in the thermodynamic limit are possible for this
simplified model. The PDSAW is used in \cite{referee13} as a model for a polymer
in the fixed-stretch ensemble. It exhibits a true phase transition from a compact
phase to a stretched phase. The critical force as a function of the temperature
can be obtained in closed form in the thermodynamic limit and does not show
reentrant behavior. In \cite{referee12}, the PDSAW is studied numerically for
finite chains. Reentrant behavior is observed in contrast to the results of
\cite{referee13} in the thermodynamic limit. It is argued in \cite{referee12}
that for the PDSAW, the value of $S_c$ is too small to cause reentrant behavior.
Reentrant behavior does show up in the numerical simulations because at small
temperatures, there is a finite entropy associated with the deformed globule.
Together with the surface term this gives rise to the observed reentrant
behavior for finite walks. Our model predicts that the reentrant behavior
survives the thermodynamic limit, in contrast to the results of the PDSAW. The
reason for this difference is that the restrictions of the PDSAW decrease the
residual entropy. 

Our model is basicly a two-state model. This kind of model has been used before
to model biopolymers, for example to study the stress-induced transformation
from B-DNA to S-DNA. In an attempt to explain the obtained experimental data,
the pure two-state model is used in \cite{referee14}. The application of the
two-state model is limited to the region of the transformation from B-DNA to
S-DNA. For this reason, the model is combined in \cite{referee15} with the well
known Worm Like Chain model. The combination of the two models results in a
reasonable fit to the experimental data. The present work uses a two-state model
in an other context, the unfolding transition of single polymers instead of the
transformation from B-DNA to S-DNA. To the best of our knowledge, this is the
first time that a two-state non-Markovian random walk is used to study the
unfolding transition of single polymer chains. We expect that a three-dimensional version of the present model can be used to study the transformation from B-DNA to S-DNA.

In conclusion, our model exhibits a gradual transition from compact globule to
rod in qualitative agreement with experimental observations. The
boundary line between these two phases shows reentrant behavior in agreement
with numerical simulations. Our model predicts that this reentrant behavior
survives the thermodynamic limit, in contrast to the results obtained for the
partially directed walk.

\section{Appendix}\label{app}
At low temperatures one can replace the average number of kinks by $\langle
K\rangle=n-\langle x\rangle/a$. With this approximation, the free energy
becomes
\begin{eqnarray}
G&=&\inf_{\langle x\rangle}\left\{h\left(n-\frac{\langle
x\rangle}{a}\right)-F\langle
x\rangle-\frac{1}{\beta}S\right\},
\end{eqnarray}
with the entropy approximately equal to
\begin{eqnarray}
\frac Sn&=&-\frac1{2na}\Big[2\langle x\rangle\ln2\langle
x\rangle+\left(na-\langle x\rangle\right)\ln\left(na-\langle
x\rangle\right)
\cr
&&-\left(na+\langle x\rangle\right)\ln\left(na+\langle
x\rangle\right)\Big].
\end{eqnarray}
The solution of the equation $\partial G/\partial\langle x\rangle=0$ gives the
following force-extension relation
\begin{eqnarray}
F&=&-\frac ha+\frac1{2a\beta}\ln\frac{4\langle x\rangle^2}{n^2a^2-\langle
x\rangle^2}.
\end{eqnarray}
After inverting this relation, one can calculate the second derivative of the
average end-to-end distance with respect to the force at constant temperature.
This second derivative equals zero if the following equation holds
\begin{eqnarray}\label{appr_Fb}
F&=&-\frac ha+\frac{\ln2}{2a}T.
\end{eqnarray}
This is a low temperature approximation for the boundary line between the
globular
and stretched phases. The factor $\ln2/2a$ is clearly a contribution due to the
entropy.

\end{document}